\documentclass{svjour3}
\usepackage{graphicx}
\begin{document}
\title {Fusion Cross Section of $\rm T(d,n)^{4}He$ and $\rm ^{3}He(d,p)^{4}He$ Reactions by Four Parameters Formula}
\author{T. Koohrokhi$^{1,}$\thanks{email:
        {\tt t.koohrokhi@gu.ac.ir}}, A. M. Izadpanah$^{1}$ and S. K. Hosseini$^{1}$}
\maketitle \centerline{\it $^{1}$Faculty of Sciences, Golestan
University, Shahid Beheshti Street, P.O. Box 155, Gorgan, IRAN}
 \vspace{15pt}
\newpage
%
%*****************************abstract**************************************
%
\begin{abstract}
Fusion cross sections of light nuclei are calculated by a complex
potential and taking into account of conservation of angular
momentum and parity. The nuclear potential is assumed to be as
simple as a spherical complex square well with a rigid core. Then,
the nuclear phase shift is extracted from continuity condition of
inverse of the logarithmic derivative of the wave functions as a
complex quantity. The quantum tunneling probability and cross
section are obtained via real and complex components of nuclear
phase shift. The obtained results for the two most important light
nuclei reactions, $\rm T(d,n)^{4}He$, $\rm ^{3}He(d,p)^{4}He$ are
compared with other theoretical formulas and experimental data.
Despite that the theory is simplified as much as possible and the
complexities and
details of nuclear interactions has been ignored, excellent agreements with experimental data are achieved.\\

 \end{abstract}
\newpage
%
%****************************section 1*******************************************
%
\section{Introduction}
A most important quantity for the analysis of nuclear reactions is
the cross section $\sigma$, which measures the probability per pair
of particles for the occurrence of the reaction. If light nuclei are
forced together, they will fuse with a yield of energy because the
mass of the combination will be less than the sum of the masses of
the individual nuclei. According to classical physics, a particle
with energy $E$ less than the height of a barrier $V_{B}$ could not
penetrate the region inside the barrier is classically forbidden.
However, quantum mechanics allows for tunneling through potential
barrier with finite range. Quantum tunneling, thus making fusion
reactions between light nuclei with energy smaller than the height
of the barrier to take place [1]. This barrier penetration effect
has important applications in various branches of nuclear physics,
specially for fusion reactions of light nuclei [2]. The fusion cross
section is proportional to the tunneling probability and a
geometrical factor $\pi \lambda^{2}\propto 1/E$, [3],
\begin{equation}
\sigma(E)\propto \frac{1}{E}\exp\left(-\frac{B_G}{\sqrt{E}}\right)
\end{equation}
where $E$ is the energy available in the center of mass (CM) frame,
$\lambda$ is the de Broglie wavelength and $B_G=\pi \alpha
Z_{p}Z_{t}e^{2}\sqrt{2\mu c^{2}}$ is the Gamow constant, expressed
in terms of the fine structure constant, $\alpha=e^{2}/\hbar c$, and
the reduced mass of the particles, $\mu=m_{p}m_{t}/(m_{p}+m_{t})$.
For completing Eq. (1), a function has been introduced which is
called astrophysical function (S-function). This function relates
nuclear part of the fusion reaction and varies slowly with energy
[4, 5]. Therefore, the fusion cross section has defined by product
of three factors,
\begin{equation}
\sigma(E)=S(E) \frac{1}{E}\exp\left(-\frac{B_G}{\sqrt{E}}\right)
\end{equation}
Definition of cross section as Eq. (2) is valid only at energies for
which the nuclear reaction is dominated by incident S-wave
transitions. Moreover, the Gamow penetrability form used here is
appropriate if the energy of the particles is much smaller than the
barrier potential [6]. To extension the validity of the cross
section expression to higher energies, a 5-parameters fitting
formula (5-P.F.) has been proposed based on the Breit-Wigner
resonance theory by Duane [7]. This formula has been widely used in
plasma fusion research [8],
\begin{equation}
\sigma(E)=\left[\frac{A_{2}}{1+(A_{3}E-A_{4})^{2}}+A_{5}\right]
\frac{1}{E\left(\exp\left(A_{1}/\sqrt{E}\right)\right)}
\end{equation}
The energy variable is the laboratory energy $E_{\rm lab}$. While
this formula accounted reasonably well for the cross section data in
the limited energy ranges of the fits, it extrapolates poorly at low
energies. Bosch and Hale improved further the formula using a
polynomial with 9 parameters based on the R-matrix theory [9],
\begin{equation}
\sigma(E)=\frac{A_{1}+E(A_{2}+E(A_{3}+E(A_{4}+EA_{5})))}{1+E(B_{1}+E(B_{2}+E(B_{3}+EB_{4})))}
\frac{1}{E}\exp\left(-\frac{B_G}{\sqrt{E}}\right)
\end{equation}
It has been shown that this 9-parameters formula (9-P.F.) yields
cross section data with much higher accuracy, leading to much better
parameterizations.

Later a 3-parameters formula has been proposed based on resonant
tunneling theory [10-12]. The latest formula contained a complex
square well potential for calculation of the S-wave ($\ell=0$) phase
shift. Although the number of parameters of this formula are less
than the 5-P.F., better results have been obtained, especially at
the low energies [13,14], but extrapolates poorly to the higher
energies.

A simple model used to account in a general way for elastic
scattering in the presence of absorptive effects is called the
optical model. In this model, the scattering is described in terms
of a complex potential. The real part, $V_{r}$, is responsible for
the elastic scattering and the imaginary part, $V_{i}$, is
responsible for the absorption. The wave number is thus complex
which follows from solving the Schrodinger equation in the usual way
for this potential. In this case, the phase shift is also complex in
general and is determined by suitable boundary conditions. In this
paper, the fusion cross section of the two most important reactions,
$\rm T(d,n)^{4}He$ and $\rm ^{3}He(d,p)^{4}He$ is calculated and
compared with other theoretical and experimental data. These
reactions are very important specially for plasma fusion [15]. The
present study is similar to 3-parameters formula that based on
resonant tunneling theory with higher partial waves and taking into
account conservation of angular momentum and parity. Furthermore, a
rigid core has been assumed for the nuclear potential to consider
the effects of Pauli exclusion and incompressibility of nuclear
matter [16]. This study proposed 4-parameters formula (4-P.F.) for
fusion cross sections that not only gets the good results for low
energies, but also extrapolates very well to higher energies.

In this paper inverse of the logarithmic derivative of the wave
functions is derived in the Sec. 2. The continuity conditions, phase
shift and cross section theory
 is reviewed in Sec. 3. Fusion cross sections of the four reactions are calculated in Secs. 4, 5 and 6.
Finally the results, including table and figures of fusion cross sections are given in Sec. 7\\

%******************************************section 2************************************

 \section{Inverse of the Logarithmic Derivative of the Wave Functions}

\subsection{Nuclear Part}

According to quantum mechanical theory for scattering, reaction
cross section is non-zero only for complex phase shifts. The real
and imaginary parts of the phase shift represent particles
scattering and absorption effects by potential, respectively. A
complex potential leads to complex phase shifts. The model contained
a complex potential is called the "optical model". Due to the short
range of nuclear force and long range of Coulomb force, the Coulomb
potential is dominant at the long distances. In the range of nuclear
force, the superiority of absorptive nuclear force compared to the
Coulomb repulsive leads to an absorptive potential well for
reaction. In the simplest case, nuclear potential well can be
considered as a spherical complex square  well with radius $R_{N}$,
and a rigid core with radius $R_{rc}$, (Fig. 1). The rigid core is
considered for including the quantum effects due to Pauli principle
and incompressibility of the nuclear matter [16],
\begin{equation}
V(r)=\left\{ \begin{array}{ll}
{\rm \infty} & {\rm if} \ r\leq R_{rc},\\
-V_{r}-\rm{i}V_{i}
& {\rm if} \ R_{rc}< r \leq R_{N},\\
\frac{Z_{p}Z_{t}e^{2}}{r} & {\rm if} \ r>R_{N}.
\end{array}
\right.
\end{equation}
where $Z_{p}e$ and $Z_{t}e$ are projectile and target charges,
respectively. According to quantum mechanics, the wave function
describing the relative motion of the two interacting nuclei
$\Psi(\textbf{r})$ is obtained by solving the Schrodinger equation.
As usual for problems characterized by a central potential, we
separate radial and angular variables, that is, we write $\Psi(r,
\theta, \phi) = Y_{\ell m}(\theta, \phi)R_{\ell}(r)$. Then the
radial part of the wave function at the range of $R_{rc}<r\leq
R_{N}$, is obtained by solving the time independent Schrodinger
equation in spherical coordinates as,
\begin{equation}
\left(\frac{d^2}{dr^{2}}+\frac{2}{r}\frac{d}{dr}\right)R_{\ell
N}(r)+\frac{2\mu}{\hbar^2}
\left(E+V_{r}+\textrm{i}V_{i}-\frac{\ell(\ell+1)\hbar^2}{2\mu
r^2}\right)R_{\ell N}(r)=0
\end{equation}
This has the familiar solutions as the regular, $j_{\ell}(\rho_{N})$
and irregular, $y_{\ell}(\rho_{N})$ spherical Bessel functions,
\begin{equation}
R_{\ell N}(\rho_{N})=Aj_{\ell}(\rho_{N})+By_{\ell}(\rho_{N})
\end{equation}
where $\rho_{N}=k_{N}r$ and $k_{N}=\sqrt{({2\mu /
\hbar^2})(E+V_{r}+\textrm{i}V_{i})}=k_{Nr}+\textrm{i}k_{Ni}$ is the
complex nuclear wave number. In Eq. 7, coefficients A and B are
determined by suitable boundary conditions. According to Eq. 5, the
nuclear potential has a rigid core so that the wave function become
zero at the point of the rigid core radius, i.e. $R_{\ell
N}(\rho_{Nc})=0$. Therefore, we have,
\begin{equation}
R_{\ell
N}(\rho_{Nc})=A\left(j_{\ell}(\rho_{N})-\frac{j_{\ell}(\rho_{Nc})}{y_{\ell}(\rho_{Nc})}y_{\ell}(\rho_{Nc}\right)
\end{equation}
where $\rho_{Nc}=k_{N}R_{rc}$. Using $R_{\ell N}(r)=u_{\ell
N}(r)/r$, the inverse of the logarithmic derivative of the nuclear
wave function at distance $r=a$ is,
\begin{equation}
\Im_{\ell N}(\eta,\rho_{a})=\frac{1}{a}\frac{u_{\ell
N}(k_{N}r)}{u_{\ell N}^\prime (k_{N}r)} \bigg \vert_
{r=a}=\frac{1}{\rho_{Na}\frac{R_{\ell N}^\prime (\rho_{Na})}{R_{\ell
N} (\rho_{Na})}+1} =N_{\ell r}(\rho_{Na})+\textrm{i}N_{\ell
i}(\rho_{Na})
\end{equation}
where $\rho_{Na}=k_{N}a$, $N_{\ell r}(\rho_{Na})$ and $N_{\ell
i}(\rho_{Na})$ are the real and imaginary parts of the inverse
logarithmic derivative of the wave function at $r=a$, respectively.

%**************************** Subsection 2.2 ************************************

\subsection{Coulomb Part}

In order to fuse, two positively charged nuclei must come into
contact, overcoming the repulsive Coulomb force. Such a situation is
made evident by the graph of the radial behavior of the potential
energy of a two nuclei system, shown in Fig. (1). The potential is
essentially Coulombian and repulsive at distances greater than
$R_{N}$. The radial part of the Coulomb wave function has the
asymptotic ($r>R_{N}$) form [17],
\begin{equation}
u_{\ell,Coul}(kr)=e^{i\delta_{\ell}}\cos \delta_{\ell}\left(\tan
\delta_{\ell}G_{\ell}(\eta,\rho)+F_{\ell}(\eta,\rho)\right)
\end{equation}
where $k=\sqrt{2 \mu E/ \hbar^2}$ is the free particle wave number,
$\eta=1 / ka_{C}$ dimensionless Coulomb parameter,
$a_{C}=\hbar^2/\mu Z_{p}Z_{t}e^2$ is Coulomb unit length,
$F_{\ell}(\eta,\rho)$ and $G_{\ell}(\eta,\rho)$ are regular and
irregular Coulomb wave functions, respectively. The inverse of the
logarithmic derivative of Coulomb wave function at $r=a$ is equal
to,
\begin{equation}
\Im_{\ell,Coul}(\eta,\rho_{a})=\frac{1}{a}\frac{u_{\ell,Coul}(kr)}{u_{\ell,Coul}^\prime
(kr)} \bigg \vert_
{r=a}=\frac{1}{\rho_{a}}\frac{F_{\ell}(\eta,\rho_{a})+\tan
\delta_{\ell}G_{\ell}(\eta,\rho_{a})}{F_{\ell}^\prime
(\eta,\rho_{a})+\tan \delta_{\ell}G_{\ell}^\prime(\eta,\rho_{a})}
\end{equation}
where $\rho_{a}=ka$.

%******************************************section 3************************************
\section{Continuity Conditions, Phase Shift and Cross Section}

The continuity conditions of wave function and its first derivative
is satisfied simultaneously by matching the nuclear and Coulomb
inverse of the logarithmic derivative of wave functions (Eq. 9 and
Eq. 11),
\begin{equation}
\Im_{\ell,\rm Coul}(\eta,\rho_{a})=\Im_{\ell,N}(\rho_{Na})
\Rightarrow \frac{1}{\rho_{a}}\frac{F_{\ell}(\eta,\rho_{a})+\tan
\delta_{\ell}G_{\ell}(\eta,\rho_{a})}{F_{\ell}^\prime
(\eta,\rho_{a})+\tan
\delta_{\ell}G_{\ell}^\prime(\eta,\rho_{a})}=N_{\ell
r}(\rho_{Na})+\textrm{i}N_{\ell i}(\rho_{Na})
\end{equation}
From this equality, phase shift is obtained as a complex quantity,
\begin{equation}
\delta_{\ell}^N=\delta_{r\ell}^N+\textrm{i}\delta_{i\ell}^N\Rightarrow\tan\delta_{\ell}^N=Td_{r\ell}^N+\textrm{i}Td_{i\ell}^N
\end{equation}
This fact follows directly from complex nuclear potential (Eq. 5).
This leads to a complex nuclear inverse of the logarithmic
derivative. The real and imaginary components of the nuclear phase
shift become,
\begin{equation}
\left \{\begin{array}{ll} \rm Td_{\ell r}^N=\frac{N_{\ell r}
\rho_{a}(G_{\ell}F_{\ell}^ \prime + F_{\ell}G_{\ell}^ \prime)
-G_{\ell}^ \prime F_{\ell}^ \prime  \rho_{a}^2 (N_{\ell r}^2+N_{\ell
i}^2)-F_{\ell}G_{\ell}}{G\prime_{\ell} ^2 \rho_{a} ^2
 (N_{\ell r}^2+N_{\ell i}^2)+G_{\ell}(G_{\ell}-2G_{\ell}^ \prime N_{\ell r}\rho_{a})} \\
\rm Td_{\ell i}^N=\frac{N_{\ell i} \rho_{a}}{G\prime_{\ell}^2
\rho_{a} ^2 (N_{\ell r}^2+N_{\ell
i}^2)+G_{\ell}(G_{\ell}-2G_{\ell}^\prime N_{\ell r}\rho_{a})}
\end{array}
\right.
\end{equation}
where in second term $(Td_{\ell i}^{N})$, Wronskian relation $
G_{\ell}F_{\ell}^ \prime - F_{\ell}G_{\ell}^\prime =1 $ is used. The
total cross section can be obtained as a sum over partial waves,
that is over the contributions of the different terms of an
expansion of the particle wave function in the components of the
angular momentum $\ell$,
\begin{equation}
\sigma_{re}=\sum_{\ell=0}^\infty \sigma_{re,\ell}
\end{equation}
the partial reaction cross section can be put in the form,
\begin{equation}
\sigma_{re,\ell}=\frac{\pi}{k^2}g(I,s_{p},s_{t})(1+\delta_{pt})P_{\ell}(E)
\end{equation}
where $\delta_{pt}$ is the Kronecker delta symbol (with
$\delta_{pt}$ = 1, if $p = t$ and $\delta_{pt}$ = 0 elsewhere) which
is introduced to properly take into account the case of reactions
between identical particles,
$g(I,s_{p},s_{t})=(2I+1)/(2s_{p}+1)(2s_{t}+1)$ is statistical factor
dependent on spin of projectile, $s_{p}$, target, $s_{t}$, and
excited state in the compound nucleus $I$. $P_{\ell}(E)$ is the
quantum-mechanical transmission probability through the potential
barrier for the $\ell-$th partial wave, i.e.
\begin{equation}
P_{\ell}(E)=\left( 1- \mid e^{2i \delta_{\ell}} \mid ^2 \right)
\end{equation}
This relation shows that transmission probability is nonzero only
for complex phase shifts. Using Eqs. 13, 14 and 17, the relationship
of the real and imaginary components of the phase shift with
transmission probability from the potential barrier is obtained,
\begin{equation}
P_{\ell}(E)=\frac{4Td_{\ell i}^N}{(Td_{\ell r}^N)^2+(1+Td_{\ell
i}^N)^2} \label{eq:16}
\end{equation}
It is essential to note that nuclear reactions follow conservation
laws. In a nuclear reaction $p+t\rightarrow C^{*} \rightarrow Y+b$,
there is conservation of total angular momentum
$\textbf{I}=\textbf{s}_{p}+\textbf{s}_{t}+\textbf{L}$ and parity
$\pi_{p} \pi_{t} (-1)^{\ell_{p,t}} = \pi(C^{*})=\pi_{b} \pi_{Y}
(-1)^{\ell_{b,Y}}$, so that these quantities must be equal on the
left and right sides of a reaction. As it can be from the following
examples, the conservation of total angular momentum and parity
limits summation of the partial waves in Eq. 15.

%*********************************** section 4 ************************************

\section{$\bf T(d,n)^{4}He$ Fusion Reaction}

Deuterium-tritium fusion reaction leads to form a compound nucleus
$\rm ^{5}He^{*}$ and then decay to $\rm ^{4}He$ and n, $\rm
T+D\rightarrow ^{5}He^{*}\rightarrow n+^{4}He$. Figure 2 shows the
nuclei rest masses and compound nucleus exited states energy levels
[18]. Spins of the deuterium and tritium nuclei are
$\textbf{s}_{D}=1$, $\textbf{s}_{T}=1/2$, respectively, and their
intrinsic parities are even. The summation of their spins is equal
to $\textbf{S}=\textbf{s}_{D}+\textbf{s}_{T}=1+1/2=\{1/2,3/2\}$.
Spin-parity of the ground state and second exited state of nucleus
$\rm ^{5}He$ is $I^{\pi}=3/2^{-}$ and $I^{\pi}=3/2^{+}$,
respectively, and statistical factor for these two states are
$g_{G.S}^{DT}=g_{2}^{DT}=2/3$. According to total angular momentum
conservation $3/2=\{1/2,3/2\}+\textbf{L}$, possible angular
momentums are $\ell=0,1,2,3$ and according to parity conservation,
for ground state $\ell=1,3$ and for second exited state $\ell=0,2$
are acceptable. Spin-parity of the first exited state of $\rm
^{5}He$ is $I^{\pi}=1/2^{-}$ and statistical factor for this state
is $g_{1}^{DT}=1/3$. According to total angular momentum
conservation, possible angular momentums are $\ell=0,1,2$ and parity
conservation required that $\ell=1$.

Now, we consider the fusion cross section of the  $\rm T(d,n)^{4}He$
reaction in the energy interval $0.12\leq E_{\rm {C.M}}(\rm
{keV})\leq 170$. According to Fig. 2, energy of the second exited
state $^{5}He$  is 137 keV higher than rest mass energy D+T. this
different energy must be compensated by the total kinetic energy of
the particles in center of mass system. Therefore, this state is
only achievable for interval energy  $137\leq E_{\rm {C.M}}(\rm
{keV})\leq 170$. For energies $0.12\leq E_{\rm {C.M}}(\rm {keV})\leq
137$ ground state and first exited state, and for energy interval
$137\leq E_{\rm {C.M}}(\rm {keV})\leq 170$ ground state, first and
second exited states are accessible. For shorthand, from then on and
in table 1, accessible energy levels are shown by $\rm EL_{a}^{b}$
notation in which a and b are the lowest and highest levels,
respectively. According to the above discussion, Eq. (16) for the
fusion cross section of the  $\rm T(d,n)^{4}He$ is,
\begin{equation}
\sigma_{f}^{DT}=\frac{\pi}{k^2}\left\{ \begin{array}{ll}
g_{G.S}^{DT} \left[ P_{1}(E)+P_{3}(E)\right] +g_{1}^{DT}
P_{1}(E)  & 0.12\leq E (\rm keV)\leq 137\\
g_{G.S}^{DT} \left[ P_{1}(E)+P_{3}(E)\right] +g_{1}^{DT}
P_{1}(E)\\
+g_{2}^{DT} \left[ P_{0}(E)+P_{2}(E)\right]  & 137\leq E (\rm keV)
\leq 170,
\end{array}
\right.
\end{equation}
The results shown in Fig. 3 by solid curve and discussed more in
section 7.

%***************************************** section 5 ************************************

\section{$\bf ^{3}He(d,p)^{4}He$ Fusion Reaction}

Deuterium-helium3 fusion reaction leads to form a compound nucleus
$\rm ^{5}Li^{*}$ and then decay to $\rm ^{4}He$ and p, $\rm
^{3}He+D\rightarrow ^{5}Li^{*}\rightarrow p+^{4}He$. Spin and parity
of the interacting particles and compound nucleus and so selection
of the angular momentums are similar to the reaction $\rm
T(d,n)^{4}He$. Fig. 4 shows the nuclei rest masses and compound
nucleus exited states energy levels [18]. Now, we consider the
fusion cross section of the $\rm ^{3}He(d,n)^{4}He$ in the energy
interval $0.54\leq E_{\rm {C.M}}(\rm {keV})\leq 850$. It is seen
that for energy range $0.54\leq E_{\rm {C.M}}(\rm {keV})\leq 484$,
ground state and first exited state ($\rm EL_{G.S}^{1}$), and for
energy interval $484\leq E_{\rm {C.M}}(\rm {keV})\leq 850$ ground
state, first and second exited states ($\rm EL_{G.S}^{2}$) are
accessible. Eq. (16) for the fusion cross section of the $\rm
^{3}He(d,n)^{4}He$ is,
\begin{equation}
\sigma_{f}^{D^3He}=\frac{\pi}{k^2}\left\{ \begin{array}{ll}
g_{G.S}^{D^3He} \left[ P_{1}(E)+P_{3}(E)\right] +g_{1}^{D^3He}
P_{1}(E)  & 0.54\leq E (\rm keV) \leq 484\\
g_{G.S}^{D^3He} \left[ P_{1}(E)+P_{3}(E)\right] +g_{1}^{D^3He}
P_{1}(E)\\
+g_{2}^{D^3He} \left[ P_{0}(E)+P_{2}(E)\right]  & 484\leq E (\rm
keV) \leq 850,
\end{array}
\right.
\end{equation}
The results shown in Fig. 5 by solid curve and discussed more in the
next section.

%*****************************section 8********************

\section{Results and Discussion}
The fusion cross sections of the two reactions, $\rm T(d,n)^{4}He$
and $\rm ^{3}He(d,p)^{4}He$ are drawn in Figs. (3) and (5)
respectively. The solid curves are based on the 4-P.F. introduced in
this paper and the results are compared with other theoretical
Formulas (5-P.F. and 9-P.F.) and experimental data. The least-square
method $S_{\rm error}=\sum_{\rm data~number}\left(\sigma_{\rm
exp}-\sigma_{\rm theory}\right)^{2}$ is applied to find the best fit
parameters and obtain values are listed in table 1. The experimental
data points are from ENDF/B VII.0 of the National Nuclear Data
Center [19]. Indeed, these points are correspond to some optimal
description of the experimental data including an averaging and
extrapolation of the available data. In spite of simplicity of the
model, the good agreements are apparent in the whole energy ranges.
The obtained values by fitting for the real, $38.5<V_{r} ~\rm
(MeV)<39.2$, and imaginary, $83<V_{i}~ \rm (keV)<187.2$, parts of
the potential indicate that $V_{i}\ll V_{r}$. Also, the ranges of
radial distances in which continuity of inverse of logarithmic
derivative of the wave functions are satisfied, are $15.2<a~(\rm
fm)<27.2$, and radius of rigid cores, are $1.2<R_{rc}~(\rm fm)<1.9$,
so that, $R_{rc}<R_{N}=1.24\left(A^{1/3}_{p}+A^{1/3}_{t}\right)<a$,
as these should be, by definition.

Although, the orders of the four parameters, $V_{r}$, $V_{i}$, $a$
and $R_{rc}$ are in accordance to the other theoretical models for
nucleus scattering [20], but these are different for different
energy ranges (Tab. 1). This is because by raising the relative
energy, more energy levels are  accessible  in the compound nucleus
and the absorption mechanism is affected by number of energy levels.
It seems the couple channels mechanism can resolve this problem
[21]. Also, It is expected that more realistic relations for the
nuclear potential improve the results. Furthermore, The effects of
some important aspects of a nuclear reaction, such as, spin-orbit
coupling, energy dependence of nuclear potential and etc, will be
analyzed in future studies. The present study is a baseline study,
which with more rich content can be also used for other reactions.
\newpage
%******************************references***************************

\newpage

%*************************Table 1*************************
\begin{table}[tbp]
\caption{Fusion cross sections experimental data.}
\begin{center}
\begin{tabular}{ |c|c|c|c|c|c|c|c|c|c|c|c| }
\hline \scriptsize{Reaction} & \scriptsize{Energy Range} &
\scriptsize{Levels} & \scriptsize{$V_{r}$
 (MeV)} & \scriptsize{$V_{i}$ (keV)} & \scriptsize{a (fm)} & \scriptsize{$R_{rc}$
 (fm)}
& \scriptsize{$S^{\rm 5-P.F.}_{\rm error}$} & \scriptsize{$S^{\rm
9-P.F.}_{\rm error}$}
& \scriptsize{$S^{\rm 4-P.F.}_{\rm error}$} \\
\hline {\scriptsize{$\rm T(d,n)^{4}He$}} &
\scriptsize{0.12$<$E$<$137} & \scriptsize{$\rm EL^{1}_{\rm G.S}$} &
\scriptsize{38.535} & \scriptsize{83}
 & \scriptsize{15.23} & \scriptsize{1.2}
& \scriptsize{0.24} & \scriptsize{0.012}
& \scriptsize{0.014}  \\
& \scriptsize{137$<$E$<$170} & \scriptsize{$\rm EL^{2}_{\rm G.S}$} &
\scriptsize{38.593} & \scriptsize{75.39}
 & \scriptsize{15.49} & \scriptsize{1.44} & & &\\
\hline {\scriptsize{$\rm ^{3}He(d,p)^{4}He$}} &
\scriptsize{0.54$<$E$<$484} & \scriptsize{$\rm EL^{1}_{\rm G.S}$} &
\scriptsize{39.076} & \scriptsize{187.12}
 & \scriptsize{27.09} & \scriptsize{1.5}
& \scriptsize{0.33} & \scriptsize{0.076}
& \scriptsize{0.005}  \\
& \scriptsize{484$<$E$<$850} & \scriptsize{$\rm EL^{2}_{\rm G.S}$} &
\scriptsize{39.195} & \scriptsize{149.94}
 & \scriptsize{27.17} & \scriptsize{1.91} & & &\\
\hline
\end{tabular}
\end{center}
\end{table}

\clearpage
%*************************fig. 1*************************
\begin{figure}
  \includegraphics[width=0.95\textwidth]{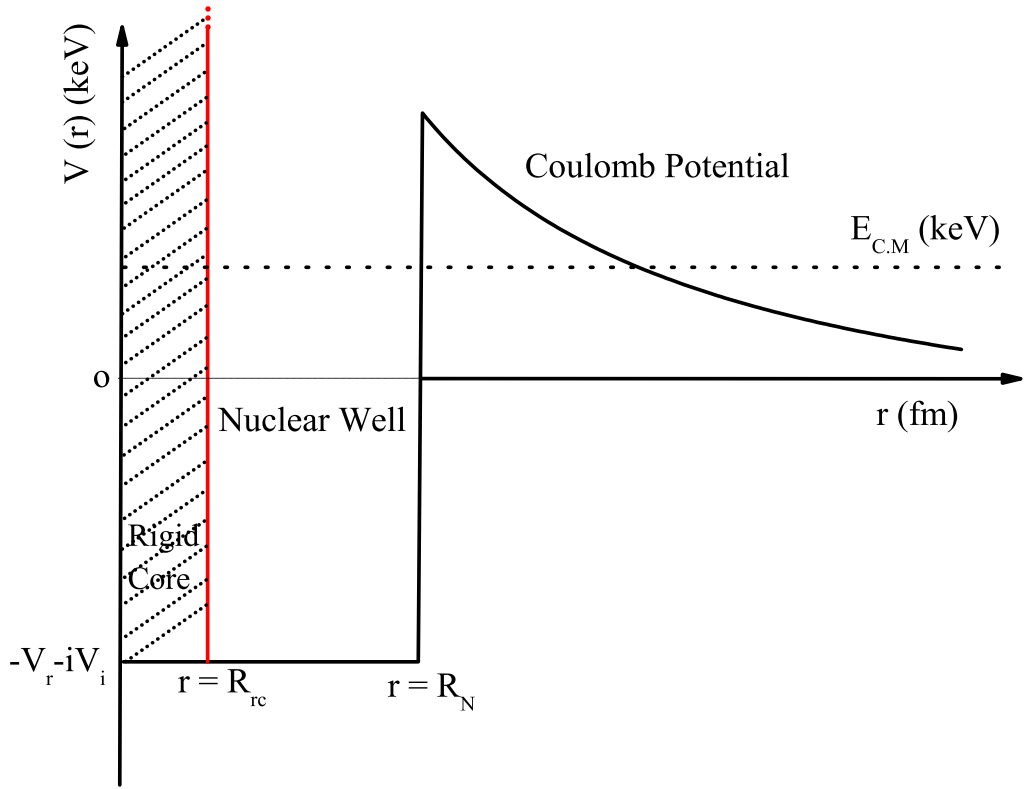}
\caption{Spherical square nuclear potential well with an imaginary
part and rigid core as well as Coulomb potential.} \label{fig:1}
\end{figure}
\clearpage
%*************************fig. 2*************************
\begin{figure}
  \includegraphics[width=0.65\textwidth]{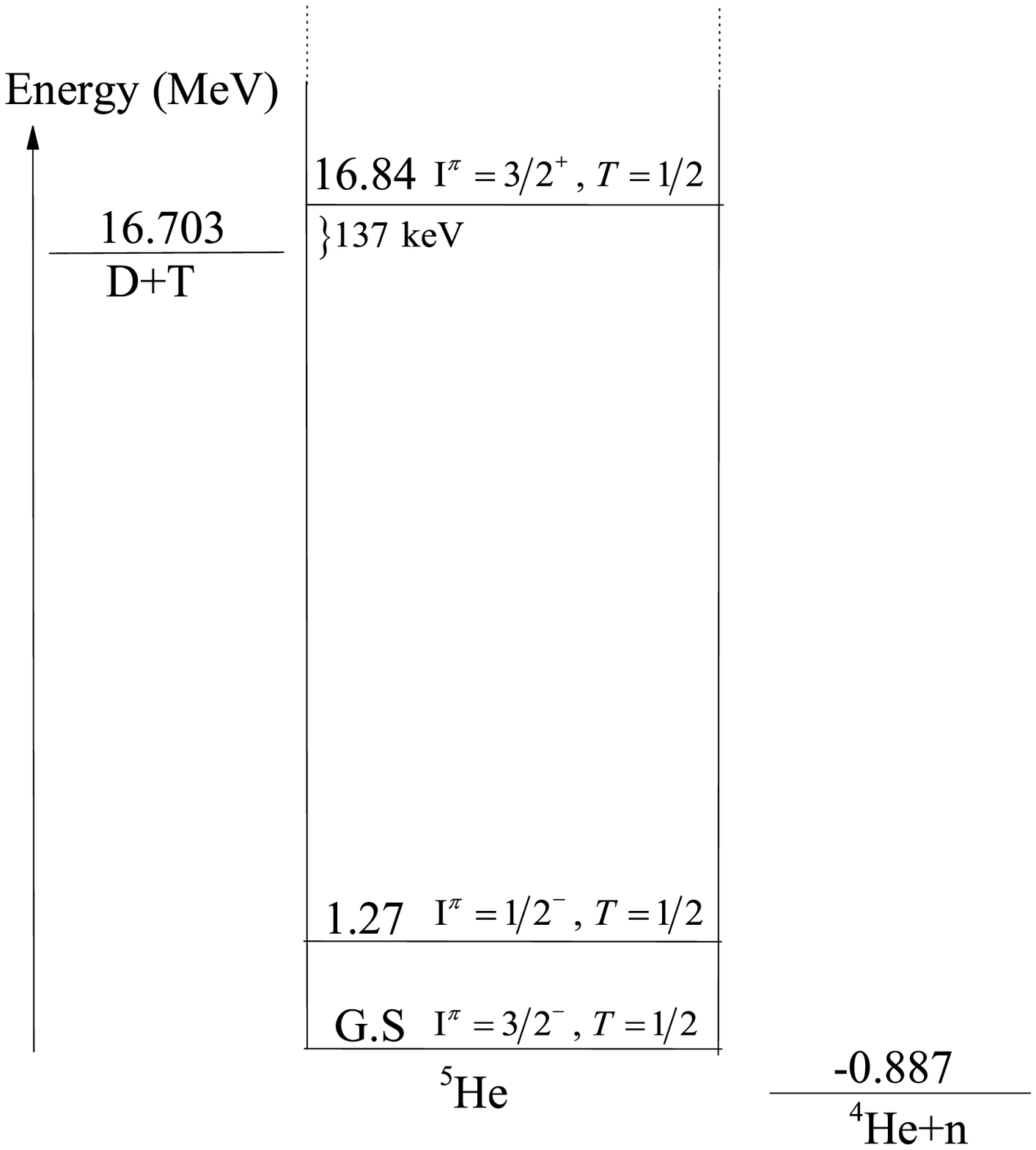}
\caption{Entrance and exit channels and $\rm
 ^{5}He$ Energy levels [18].}
\label{fig:2}
\end{figure}
\clearpage
%*************************fig. 3*************************
\begin{figure}
  \includegraphics[width=0.95\textwidth]{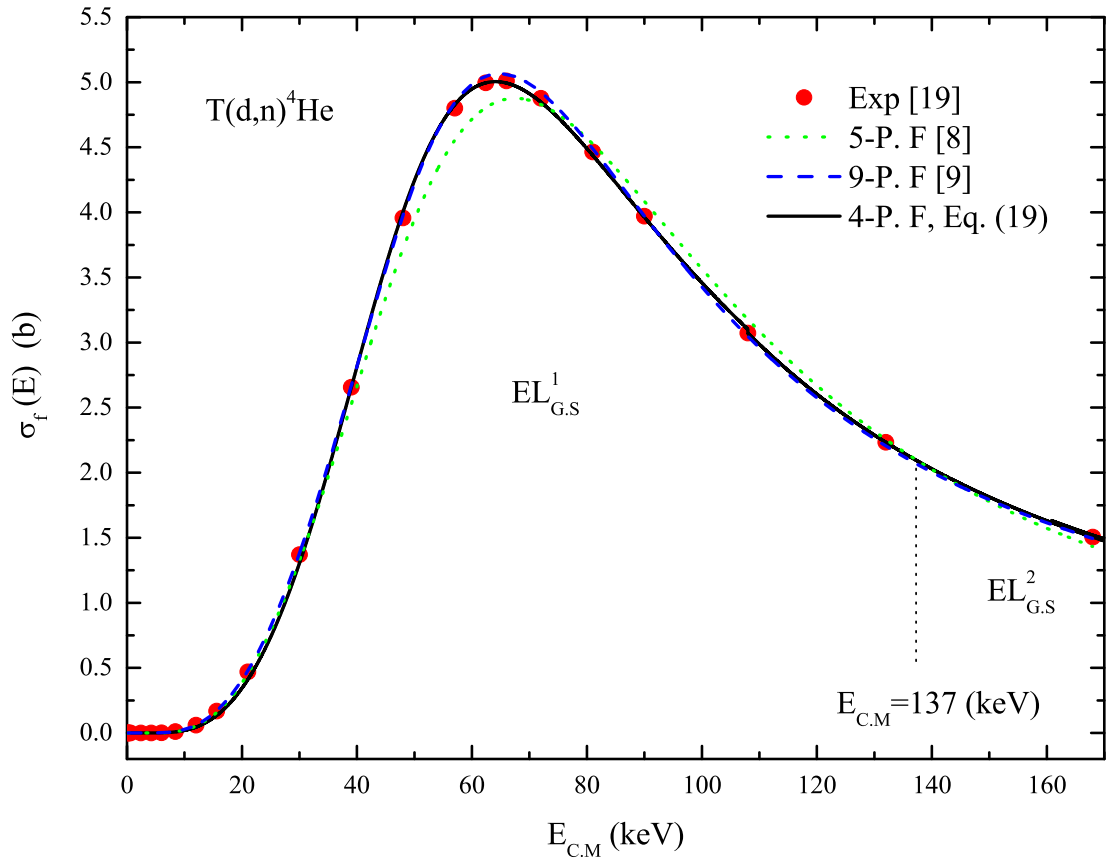}
\caption{Comparison between fusion data and the fusion cross
sections of $\rm T(d,n)^{4}He$ calculated by 5-P.F, 9-P.F and
4-P.F.} \label{fig:3}
\end{figure}
\clearpage
%*************************fig. 4*************************
\begin{figure}
  \includegraphics[width=0.65\textwidth]{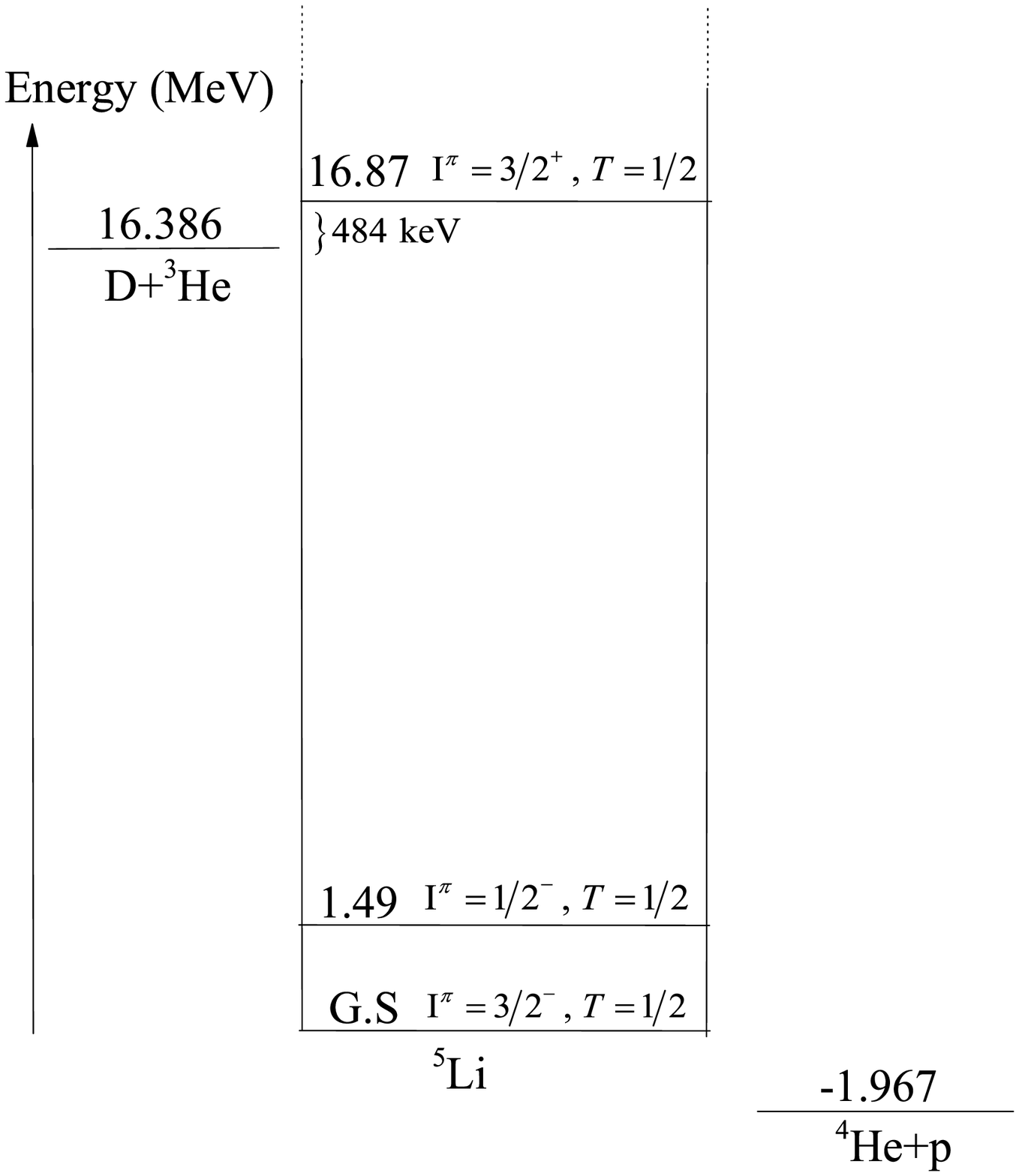}
\caption{Entrance and exit channels and $\rm ^{5}Li$ Energy levels
[18].} \label{fig:4}
\end{figure}
\clearpage
%*************************fig. 5*************************
\begin{figure}
  \includegraphics[width=0.95\textwidth]{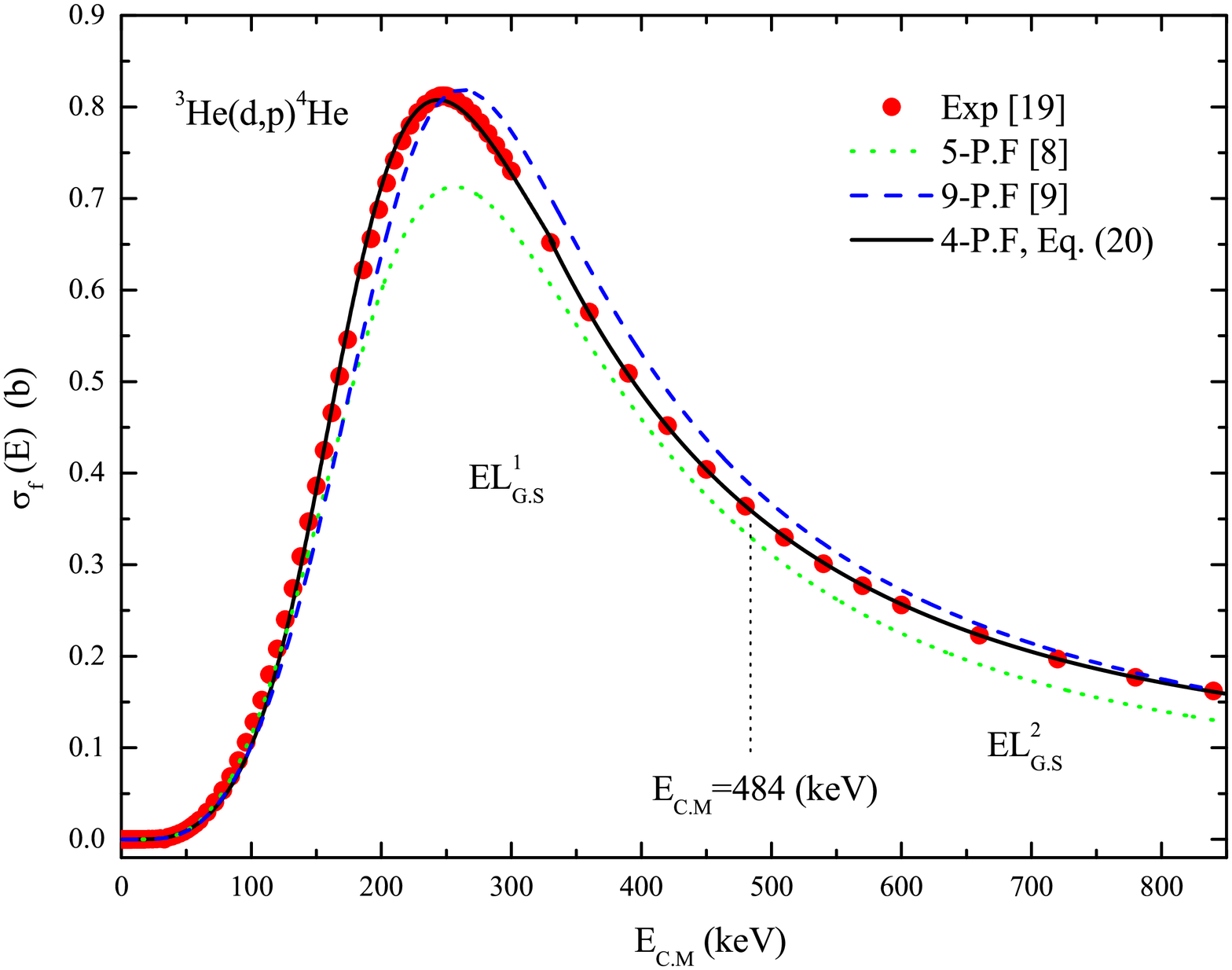}
\caption{Comparison between fusion data and the fusion cross
sections of $\rm ^{3}He(d,p)\rm ^{4}He$ calculated by 5-P.F, 9-P.F
and 4-P.F.} \label{fig:5}
\end{figure}
\clearpage
\end{document}